\documentclass[aps,reprint,superscriptaddress,article]{revtex4-2}
\usepackage{graphicx,graphics,color,epsfig}
\usepackage{times}
\usepackage{epstopdf}
\usepackage{bm}
\usepackage{amsmath}
\usepackage{amssymb}
\usepackage{diagbox}
\usepackage{array}
\usepackage{color}
\usepackage{float}
\usepackage{appendix}
\usepackage{dcolumn}
\usepackage[colorlinks=true, letterpaper=true, pdfstartview=FitV, linkcolor=blue, citecolor=blue, urlcolor=blue]{hyperref}

\begin{document}

\preprint{APS/123-QED}

\title{Pressure and doping control of magnetic order and metallization in Ruddlesden-Popper La$_2$NiO$_4$}

\author{Han-Yu Wang}\thanks{These authors contributed equally to this work.}
\affiliation{Key Laboratory of Materials Physics, Institute of Solid State Physics, HFIPS, Chinese Academy of Science, Hefei 230031, China}
\affiliation{Science Island Branch of the Graduate School, University of Science and Technology of China, Hefei 230026, China}

\author{Shu-Hong Tang}\thanks{These authors contributed equally to this work.}
\affiliation{Key Laboratory of Materials Physics, Institute of Solid State Physics, HFIPS, Chinese Academy of Science, Hefei 230031, China}
\affiliation{Science Island Branch of the Graduate School, University of Science and Technology of China, Hefei 230026, China}

\author{Xiao-Teng Huang}
\affiliation{Key Laboratory of Materials Physics, Institute of Solid State Physics, HFIPS, Chinese Academy of Science, Hefei 230031, China}
\affiliation{Science Island Branch of the Graduate School, University of Science and Technology of China, Hefei 230026, China}

\author{Ya-Min Quan}
\affiliation{Key Laboratory of Materials Physics, Institute of Solid State Physics, HFIPS, Chinese Academy of Science, Hefei 230031, China}

\author{Xian-Long Wang}
\affiliation{Key Laboratory of Materials Physics, Institute of Solid State Physics, HFIPS, Chinese Academy of Science, Hefei 230031, China}
\affiliation{Science Island Branch of the Graduate School, University of Science and Technology of China, Hefei 230026, China}

\author{Yan-Ling Li}
\affiliation{School of Physics and Electronic Engineering, Jiangsu Normal University, Xuzhou 221116, China}

\author{Da-Yong Liu}
\affiliation{School of Physical Science and Technology, Nantong University, Nantong 226019, China}

\author{H.-Q. Lin}
\affiliation{School of Physics, Zhejiang University, Hangzhou 310058, China}

\author{Zhi Zeng}
\affiliation{Key Laboratory of Materials Physics, Institute of Solid State Physics, HFIPS, Chinese Academy of Science, Hefei 230031, China}
\affiliation{Science Island Branch of the Graduate School, University of Science and Technology of China, Hefei 230026, China}

\author{Liang-Jian Zou}
\email{zou@theory.issp.ac.cn}
\affiliation{Key Laboratory of Materials Physics, Institute of Solid State Physics, HFIPS, Chinese Academy of Science, Hefei 230031, China}
\affiliation{Science Island Branch of the Graduate School, University of Science and Technology of China, Hefei 230026, China}

\begin{abstract}
  The discovery of superconductivity in multilayer nickelates under pressure has intensified interest in understanding the magnetic and electronic properties of Ruddlesden-Popper nickelates. Using density functional theory with Hubbard corrections, we investigate the magnetic ground state, electronic structure evolution under pressure, and Sr-doping effects in La$_2$NiO$_4$. We find that at ambient pressure, tetragonal La$_2$NiO$_4$ exhibits G-type antiferromagnetic order with negligible interlayer magnetic coupling. Under hydrostatic pressure, the system undergoes a continuous insulator-metal transition at ~50 GPa while maintaining robust magnetic order up to 75 GPa, contrasting sharply with the rapid magnetic suppression in La$_3$Ni$_2$O$_7$. Sr doping induces a systematic evolution from G-type to A-type, to striped antiferromagnetic orders, and eventually to ferromagnetic order, accompanied by metallization. Furthermore, LaSrNiO$_4$ displays weak charge and orbital orders. These results reveal the unique pressure and doping effects of single-layer nickelates and provide insights into the magnetic mechanisms underlying nickelate superconductivity.
\end{abstract}

\maketitle

{\bf Motivations:}
The recent discovery of  superconductivity in Ruddlesden-Popper (RP) nickelates~\cite{li2019superconductivity, wang2022pressure, sun2023signatures, Zhu2024Nature, Zhang2025Superconductivity, Zhang2025General} under high pressure provides a new platform for exploring unconventional high-temperature superconductivity.
Among the RP series La$_{n+1}$Ni$_n$O$_{3n+1}$, the single-layer compound La$_2$NiO$_4$ ($n=1$) is the most fundamental member, it may provide the understand of the intrinsic electronic and magnetic nature of the NiO$_2$ planes. Although synthesized shortly after the discovery of the cuprate superconductor La$_{2-x}$Sr$_x$CuO$_4$ \cite{bednorz1986possible, cava1987bulk}, the parent phase La$_2$NiO$_4$ and its Sr-doped variants have not exhibited superconductivity at ambient pressure \cite{Cava1991Magnetic,Eisaki1992Eisaki}.
This raises a fundamental question: what are the electronic and magnetic properties of La$_2$NiO$_4$? Why does La$_2$NiO$_4$, unlike its cuprate analogue La$_{2-x}$Sr$_x$CuO$_4$, not exhibit superconductivity at ambient pressure?
Furthermore, are these physical properties analogous to or fundamentally different from those observed in La$_3$Ni$_2$O$_7$ and La$_4$Ni$_3$O$_{10}$?
These questions underscore the necessity for a comprehensive and systematic investigation of La$_2$NiO$_4$.

It is important to theoretically reveal the magnetic phase diagram of La$_{2}$NiO$_4$ under hydrostatic pressure, as well as the dependence of doping concentration. From the view of the superconducting pairing mechanism, it is crucial to reveal the position and intensity of spin resonance peaks in the spin fluctuation spectra of the theoretical model available and the comparison with experimental observation \cite{monthoux1991towards}. However, up to date, most of the researches assumed that the coupling between Ni spins is a Neel-type antiferromagnetic interaction in La$_2$NiO$_4$, as in La$_2$CuO$_4$ \cite{Petsch2023High}. Drawing on insights from iron-based superconductors, one may expect that the multi-orbital nature and partial orbital filling in the high-pressure phases of RP nickelates give rise to magnetic interactions and ground states that differ from the simple Neel antiferromagnetic order of single-orbital cuprates \cite{Veitti2025Superconductivity,Zhang2024Electronic}. Such complexity is expected to profoundly affect the magnetic order of the parent phase, the spin-fluctuation spectrum, the pairing symmetry, and the possible emergence of a spin-resonance mode.
It is worth investigating whether tetragonal La$_{2-x}$A$_x$NiO$_4$ (A=Ca, Sr, ...) is or is not superconducting under high pressure, and also worth carrying out extensively and systematically exploring the similar or different properties of La$_{2-x}$A$_x$NiO$_4$ with respect to La$_3$Ni$_2$O$_7$ and La$_4$Ni$_3$O$_{10}$ nickelates.

In this \textit{Paper}, by performing Hubbard corrections (DFT+$U$) calculations using the Vienna Ab initio Simulation Package (VASP) with projector-augmented wave pseudopotentials, we systematically map out the magnetic and electronic phase diagram of La$_2$NiO$_4$ as a function of pressure and Sr doping.
We apply the hydrostatic pressure from ambient up to 100 GPa and ensure structural stability. For Sr-doped compounds La$_{2-x}$Sr$_x$NiO$_4$ ($x=0.5,1.0,1.5$), we use ordered substitution patterns and perform structural relaxations to ensure the thermodynamic stability.
The Hubbard $U$ parameter for Ni 3$d$ orbitals was varied from 2 eV to 5 eV, with Hund's coupling $J = 0.1U$.
We demonstrate that while the ambient-pressure ground state is a G-type antiferromagnet, this state is remarkably sensitive to doping, while less to pressure. We reveal a pressure-induced insulator-to-metal transition around 50 GPa. Furthermore, we uncover a rich magnetic phase diagram with Sr doping, where the ground state evolves from G-type to A-type antiferromagnetic, to a striped spin-charge-orbital order, and ultimately to ferromagnetic order. Our findings provide a benchmark for understanding the fundamental properties of the 214 nickelate and contribute to understanding the interplay of magnetism and superconductivity across the RP nickelate family.

{\bf Magnetic  Groundstate at Ambient Pressure:}
\begin{figure*}[htbp]
  \includegraphics[width=1\linewidth]{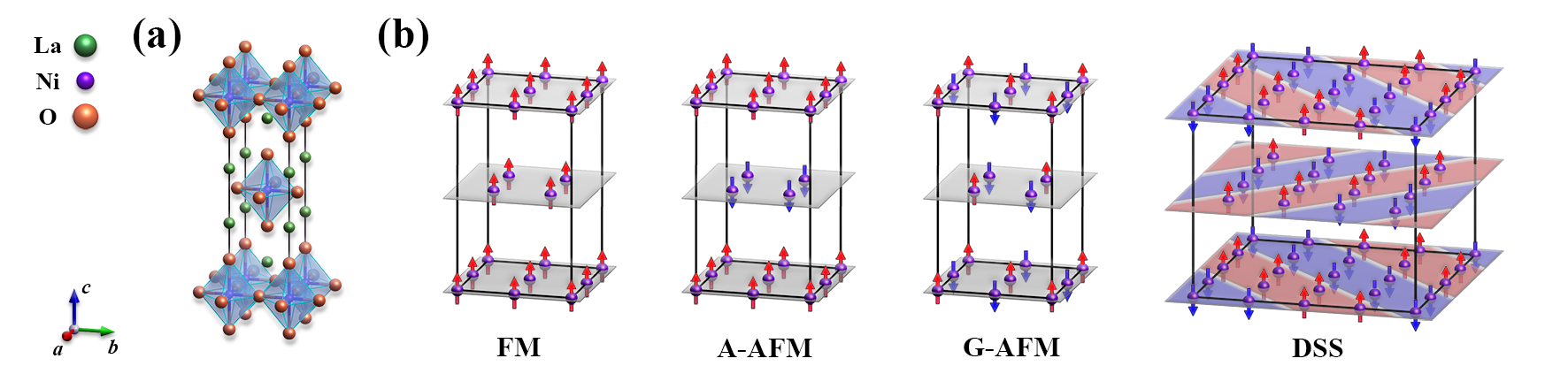}
  \caption{\label{fig1} Crystal structure and schematic of candidate magnetic configurations for La$_2$NiO$_4$. Red and blue arrows denote Ni-site spin-up and spin-down orientations, respectively. The considered magnetic orders include ferromagnetic (FM), A-type antiferromagnetic (A-AFM), G-type antiferromagnetic (G-AFM), and double spin stripe (DSS) configurations.}
\end{figure*}
We begin by determining the magnetic ground state of La$_2$NiO$_4$ at ambient pressure using density functional theory with Hubbard $U$. To systematically evaluate competing magnetic configurations, we consider five distinct spin arrangements: (1) non-magnetic (NM), and four ordered states illustrated in Fig.~\ref{fig1}, namely (2) ferromagnetic (FM), (3) A-type antiferromagnetic (A-AFM), (4) G-type antiferromagnetic (G-AFM), and (5) double spin stripe (DSS) configurations.
Our calculations span a range of Hubbard $U$ values from 2 to 5 eV, allowing us to assess the impact of electron correlations on the magnetic stability. The Hund's exchange coupling is held at a conventional value of $J_H = 0.1U$, consistent with prior studies on nickelates.

\begin{figure*}[htbp]
  \includegraphics[width=1\linewidth]{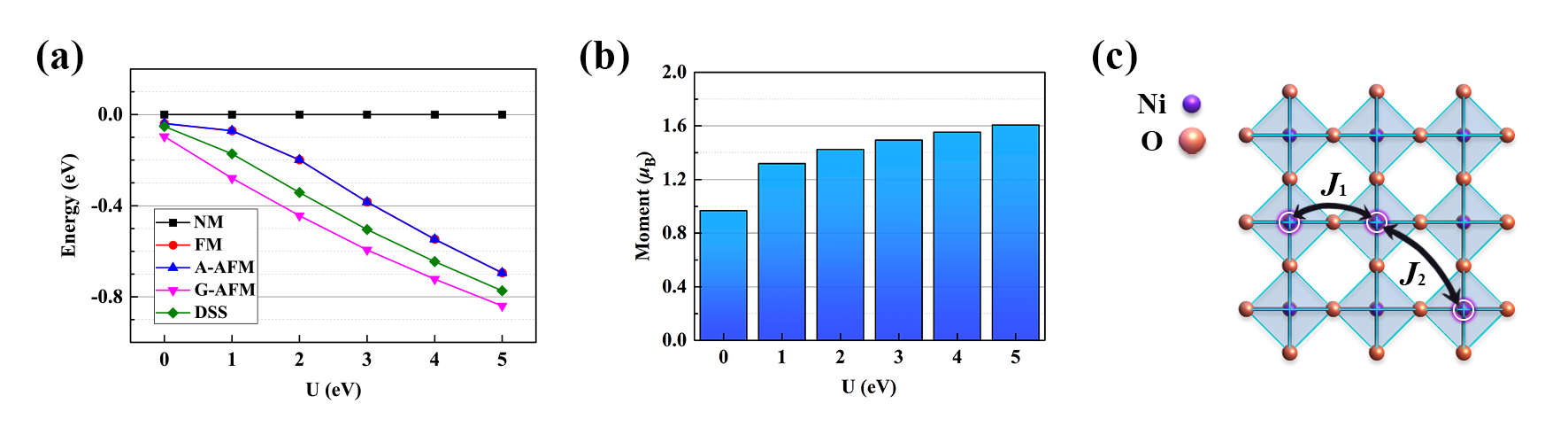}
  \caption{\label{fig2} Magnetic stability and exchange interactions in La$_2$NiO$_4$. (a) Total energy differences of various magnetic configurations relative to the NM state as a function of the Hubbard $U$ parameter. (b) Calculated local magnetic moments of Ni ions across the same $U$ range. (c) Schematic illustration of the nearest-neighbor $J_{1}$ and next-nearest-neighbor $J_{2}$ exchange coupling paths within the NiO$_2$ planes.}
\end{figure*}

Fig.~\ref{fig2}(a) presents the total energy differences per conventional unit cell (La$_4$Ni$_2$O$_8$) across all considered magnetic states. For each $U$ value, the total energy is normalized relative to the NM configuration, enabling direct comparison of magnetic phase stability. All magnetic configurations are energetically favored over the NM state, underscoring the crucial role of electron correlations. Most notably, the G-AFM state consistently exhibits the lowest energy across the entire $U$ range, firmly establishing it as the magnetic ground state. The energy difference between G-AFM and the next-lowest DSS configuration is approximately 0.1~eV, demonstrating the robust stability of the G-AFM phase. The near-degeneracy between FM and A-AFM energies indicates negligible interlayer magnetic coupling, which is consistent with the quasi-two-dimensional nature of the material.
Meanwhile, Fig.~\ref{fig2}(b) displays the calculated local magnetic moments of Ni ions for all the magnetic configurations, which exhibit values in the range of 1.40-1.75 $\mu_{B}$ depending on the specific configuration and Hubbard $U$ value. Notably, the moment sizes exhibit only weak dependence on the Hubbard $U$ parameter, suggesting robust local moment formation regardless of the precise correlation strength.

{\bf Magnetic interactions at ambient pressure:}
To further elucidate the magnetic behavior of parent La$_2$NiO$_4$, we analyze the spin interactions in insulating A-AFM phase by using the Heisenberg model:
\begin{eqnarray}
 \mathcal{H} = \sum_{i<j} J_{ij} \mathbf{S}_i \cdot \mathbf{S}_j
\end{eqnarray}
where $J_{ij}$ represents the exchange coupling between Ni spins $\mathbf{S}_i$ and $\mathbf{S}_j$, with the spin quantum number $S=1$ corresponding to the 3$d^{8}$ electronic configuration of Ni$^{2+}$ in the low-spin state.
The exchange parameters are extracted by mapping the DFT+U energies of various collinear magnetic states to the Heisenberg model through the four-state method.

\begin{table}[htbp]
  \setlength{\belowcaptionskip}{15pt}
  \caption{\label{T1} The Heisenberg exchange interactions of nearest-neighbor $J_{1}$ and next-nearest-neighbor $J_{2}$ with different values of Hubbard $U$.}
  \renewcommand{\arraystretch}{1.2}
  \begin{ruledtabular}
  \begin{tabular}{m{2cm}<{\centering}|cc|cc}
  $U$ (eV)  &  $J_{1}$ (meV) & &  $J_{2}$ (meV)    \\ \hline
  2         &      61.2     & &     2.6          \\
  3         &      52.8     & &     2.8          \\
  4         &      43.8     & &     2.1          \\
  5         &      36.2     & &     1.5          \\
  \end{tabular}
  \end{ruledtabular}
\end{table}
In parent phase La$_2$NiO$_4$, the shortest interlayer Ni--Ni distance is 6.8~\AA, and the near-degeneracy between FM and A-AFM energies justifies neglecting the interlayer exchange $J_{\perp}$. For in-plane couplings, we restrict our analysis to Ni-Ni distances of 6.8 \AA\, retaining only nearest-neighbor $J_{1}$ and next-nearest-neighbor $J_{2}$ interactions within the NiO$_2$ planes, as illustrated in Fig. \ref{fig2}(c). The calculated $J_1$ and $J_2$ values for $U = 2$--5~eV are summarized in Table~\ref{T1}.
The nearest-neighbor exchange coupling $J_1$ is consistently positive, ranging from 36.2 meV to 61.2~meV, indicating robust antiferromagnetic coupling. Its systematic decrease with increasing $U$ reflects the suppression of charge fluctuations due to enhanced electron correlation. In contrast, $J_2$ remains weak (1.5-2.8~meV) and exhibits minimal $U$-dependence, suggesting its origin may stem from direct exchange or long-range hybridization rather than on-site Coulomb effects.

The small $J_2/J_1$ ratio (0.04-0.05) accounts for the stability of the G-AFM ground state observed in DFT+$U$ calculations, as it is insufficient to induce magnetic frustration that could stabilize competing configurations such as the DSS phase. Notably, the magnitude of $J_1$ agrees well with experimental estimates from similar nickelates~\cite{bialo2024strain}. The gradual reduction of $J_1$ with increasing $U$ suggests that intermediate correlation strengths yield the best agreement with experiments.
Overall, these results demonstrate how the strength and $U$-dependence of exchange interactions govern the magnetic properties of La$_2$NiO$_4$. The dominant $J_1$ coupling dictates the robust AFM character, while the weak $J_2$ term has negligible influence on ground-state selection. The quantitative agreement between theoretical predictions and experimental observations validates our computational approach and provides microscopic insight into the exchange mechanisms in this strongly correlated oxide system.

{\bf Pressure Evolution of Electronic Structures:}
We further investigate the magnetic ground state and electronic structure of La$_2$NiO$_4$ under varying hydrostatic pressures. Pressure serves as a clean tuning parameter that can modify electronic bandwidth, orbital overlap, and magnetic exchange interactions without introducing chemical disorder. The results show that the system consistently retains an intralayer AFM configuration with negligible interlayer spin coupling. Both crystallographically distinct Ni atoms exhibit identical magnetic moments of 1.6~$\mu_B$ at ambient pressure, which gradually decrease to approximately 1.4~$\mu_B$ at 75~GPa. The absence of magnetic moment splitting indicates a uniform spin-density-wave phase without any sign of charge or orbital ordering in pressured parent phase La$_2$NiO$_4$ up to 100~GPa.

\begin{figure*}[htbp]
  \includegraphics[width=1\linewidth]{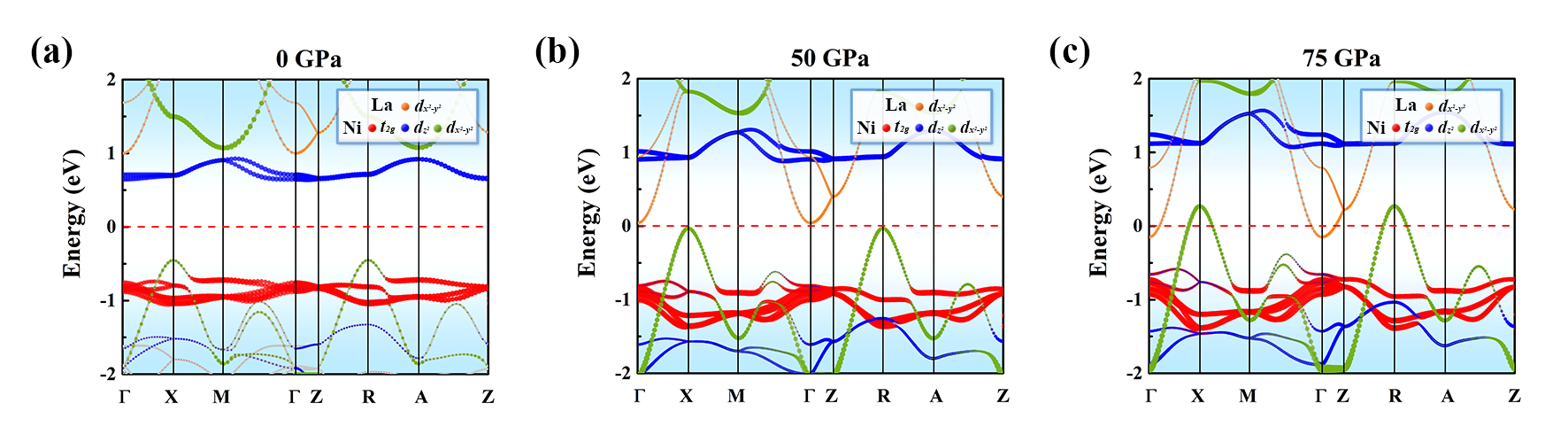}
  \caption{\label{fig3} Pressure-induced evolution of the electronic structure in G-AFM La$_2$NiO$_4$. Band structures are shown for hydrostatic pressures of (a) 0 GPa, (b) 50 GPa, and (c) 75 GPa. The progression illustrates a continuous insulator-to-metal transition with increasing pressure.}
\end{figure*}

Fig.~\ref{fig3}(a--c) presents the pressure-dependent band structures of La$_2$NiO$_4$ in its magnetic ground state. From 0~GPa to 50~GPa, the system remains insulating, with a finite energy gap near the Fermi level. This gap, initially about 1~eV under ambient conditions, gradually narrows with increasing pressure and eventually closes at around 50~GPa. The dynamical of the high-pressure $I4/mmm$ phase is confirmed by our phonon dispersion calculations, see Section V in the Supplemental Material for details. At higher pressures (50--100~GPa), La$_2$NiO$_4$ enters a metallic state. This continuous gap evolution suggests a smooth insulator-to-metal transition without abrupt structural phase transitions, consistent with experimental observations~\cite{Ji2024-CPL}.

Our results are further corroborated by partial density of states (PDOS) analyses. The orbital-resolved density of states (DOS) of the two inequivalent Ni sites at 75~GPa display identical electronic occupations despite opposite spin orientations. This indicates spatially uniform charge and orbital distributions. Across all pressures, the Ni atoms maintain the same electronic configuration: ($t_{2g}^6 d_{z^2}^1 d_{x^2-y^2}^1$), corresponding to a Ni$^{2+}$ valence with fully occupied $t_{2g}$ orbitals and half-filled $e_g$ orbitals. Although the distorted octahedral crystal field lifts the degeneracy between the $d_{z^2}$ and $d_{x^2-y^2}$ orbitals, strong on-site Coulomb repulsion prevents double occupancy and favors electron localization. As a result, orbital polarization remains absent, and orbital symmetry is preserved despite symmetry-breaking lattice distortions. This highlights the dominant role of electronic correlations in stabilizing the orbital hierarchy and maintaining the insulating phase.

In contrast to the DSS configuration observed in La$_3$Ni$_2$O$_7$~\cite{Zhao2025-ScienceBulletin,Ni2025-NPJ,Zhang2024-arxiv,LaBollita2024-PRM}, La$_2$NiO$_4$ does not display evidence of charge or orbital ordering. The distorted crystal field that would typically drive orbital differentiation is effectively counteracted by strong electronic correlations, as represented by the Hubbard $U$ parameter. This compensation preserves orbital uniformity even under symmetry-breaking lattice distortions, positioning La$_2$NiO$_4$ uniquely in the interplay between lattice geometry and electron correlation.
Importantly, the insulating gap is absent in nonmagnetic configurations, even with substantial Hubbard $U$, indicating that the insulating gap is stabilized by long-range magnetic order, consistent with a Slater-like or correlation-assisted Slater insulating scenario. Compared to its RP counterpart La$_3$Ni$_2$O$_7$, La$_2$NiO$_4$ displays fundamentally distinct pressure responses. While both systems undergo metallization under pressure, La$_3$Ni$_2$O$_7$ shows rapid suppression of spin-charge-orbital ordering with the onset of metallicity near 10~GPa and a sharp reduction of Ni magnetic moments from 1~$\mu_B$ to 0.6~$\mu_B$ at 40~GPa~\cite{Zhao2025-ScienceBulletin,Zhang2024-arxiv,Khasanov2024-arxiv,Guo1988-JPCS,sun2023signatures}, coinciding with the emergence of 80~K superconductivity.

By contrast, La$_2$NiO$_4$ exhibits exceptional magnetic robustness: Ni moments remain at 1.4~$\mu_B$ even under 75~GPa compression, and full metallization only occurs above 50~GPa.
In La$_3$Ni$_2$O$_7$, pressure promotes interlayer $d_{z^2}$ orbital hybridization, which weakens magnetic interactions. Conversely, monolayer La$_2$NiO$_4$ retains dominant in-plane $d_{x^2 - y^2}$ orbital character, protected by pressure-resistant superexchange mechanisms. This persistent magnetism suggests that superconductivity is unlikely to emerge in La$_2$NiO$_4$ before  hydrostatic pressure reaches 50~GPa. Nevertheless, chemical strategies, such as interlayer engineering via Sr substitution~\cite{Zhang2024-JMS,Kumar2021-MaterialsToday} may help suppress magnetism while preserving correlation effects, potentially offering alternative routes toward high-$T_\mathrm{c}$ superconductivity in monolayer nickelates, as addressed in what follows. Moreover, we note that our conclusions are based on static DFT+$U$ calculations with collinear magnetic order; possible subtle orbital instabilities driven by dynamic correlations are beyond the present scope.

{\bf Effects of Sr-doping on La$_2$NiO$_4$:}
Given the robust magnetism of La$_2$NiO$_4$ under pressure, we explore an alternative approach AFM insulating behavior, with its electronic structure near the Fermi level dominated by Ni $3d^{8}$ states, particularly in the $d_{x^{2}-y^{2}}$ orbital, under high pressure. Previous studies suggest that Sr doping can induce an insulator-to-metal transition in this system~\cite{Zhang2024-JMS}. To further understand the role of doping, we investigate the electronic and magnetic properties of La$_{2-x}$Sr$_x$NiO$_4$ for doping levels $x=0.5,1.0,1.5$. The optimized crystal structures for various doping levels are provided in Sections II of the Supplemental Material. The partial substitution of Sr$^{2+}$ for La$^{3+}$ introduces both chemical pressure and hole carriers, which are expected to simultaneously trigger lattice distortions and charge reconstruction. These changes can significantly influence the insulator-metal transition and alter Ni-O bond covalency.

\begin{figure*}[htbp]
  \includegraphics[width=0.8\linewidth]{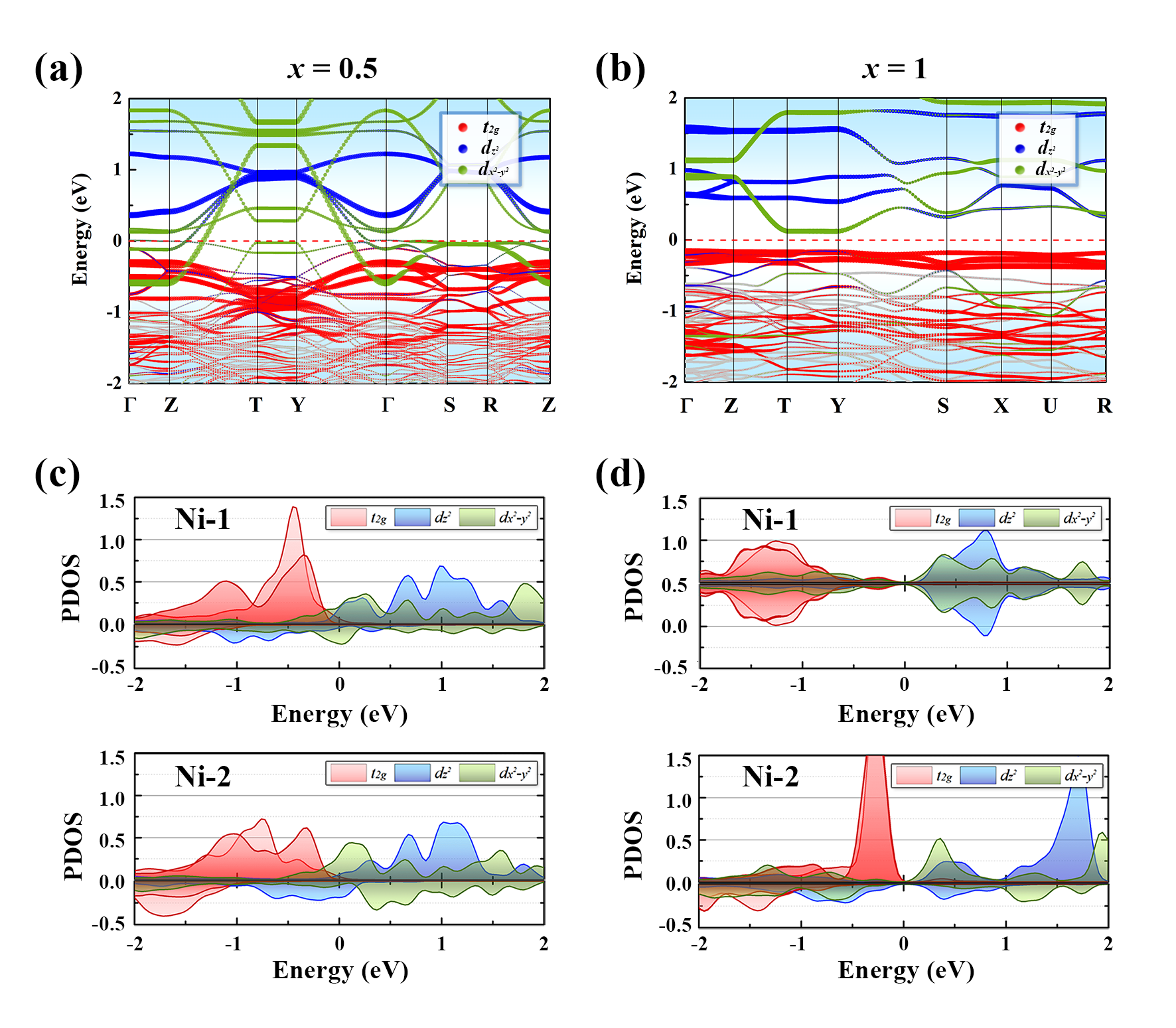}
  \caption{\label{fig4} Doping-dependent electronic properties of La$_{2-x}$Sr$_x$NiO$_4$. Band structures are presented for (a) $x$=0.5 and (b) $x$=1.0. Panels (c) and (d) show the corresponding spin-resolved and orbital-projected partial density of states (PDOS) for the two crystallographically inequivalent Ni sites, highlighting the impact of hole doping on orbital occupancy and site symmetry.}
\end{figure*}

\begin{figure}[htbp]
  \includegraphics[width=0.8\linewidth]{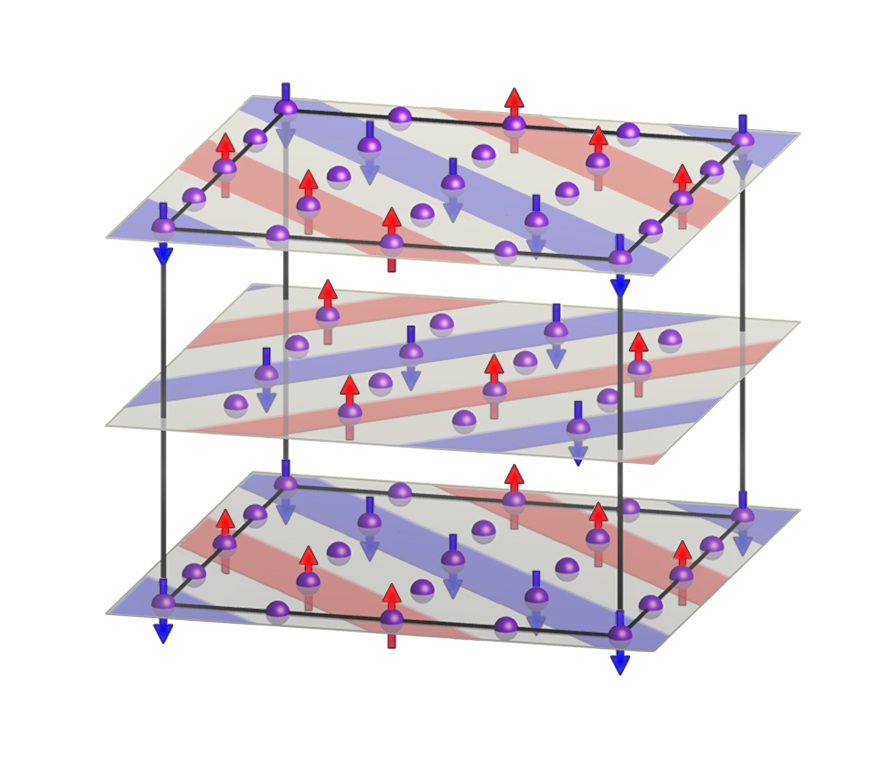}
  \caption{\label{fig5} Predicted ground-state magnetic configuration for Sr-doped La$_{2-x}$Sr$_x$NiO$_4$ at $x=1$.}
\end{figure}

In all cases, the ordered substitution of La by Sr atoms breaks the original tetragonal symmetry of the parent compound ($I4/mmm$ space group), leading to a symmetry-lowering structural distortion. As a consequence, Ni sites split into two inequivalent types: one preserves a near-ideal NiO$_2$ planar geometry, while the other experiences local distortions driven by adjacent Sr dopants.
Our calculations reveal the magnetic ground states and a doping-induced sequence of magnetic transitions. Specifically, at $x=0.5,1.0,1.5$ the compounds stabilize in A-AFM, striped AFM [Fig.~\ref{fig5}], and FM configurations, respectively, distinct from the G-AFM order in the undoped parent phase. This evolution of magnetic order stems from a systematic reconstruction of the Ni $e_g$ orbital occupancy due to hole doping. The substitution of Sr$^{2+}$ for La$^{3+}$ introduces holes that preferentially affect the $d_{x^2-y^2}$ orbital.

As illustrated in Fig.~\ref{fig4}(a, c), the electronic structures for $x=0.5$ exhibits metallic characteristics, with multiple bands crossing the Fermi level. This behavior indicates the collapse of the insulating gap and the emergence of a correlated multiband metallic state. The primary contributions to the DOS near the Fermi level arise from hybridized Ni $3d$ and O $2p$ orbitals, a characteristic feature that mirrors the pressure-driven metallization observed in the undoped compound.
In contrast, the $x=1.0$ system exhibits insulating behavior and striped spin order, as evidenced in the spin-resolved band structure [Fig.~\ref{fig4}(b)] and partial DOS [Fig.~\ref{fig4}(d)]. Insulating LaSrNiO$_4$ opens a gap of 0.27~eV, and spins form an stripe with alternative $S=0$ and $S=1$ along the (110) direction. The partial DOS reveals a pronounced disparity between the two inequivalent Ni sites, indicating charge and orbital disproportionations. This spin-charge-orbital ordering likely is consistent with a site-selective Mott-like scenario or Coulomb-driven symmetry breaking \cite{Greenberg2018Pressure}, where the localization occurs preferentially on specific Ni sites.

Doping also leads to progressive suppression of local magnetic moments. For $x=0.5$, the two inequivalent Ni sites host moments of 1.5~$\mu_B$ and 1.2~$\mu_B$, respectively. These values further decrease to 1.5~$\mu_B$ and 0.0~$\mu_B$ at $x=1.0$, and to 0.9~$\mu_B$ and 0.8~$\mu_B$ at $x=1.5$, significantly lower than the 1.6~$\mu_B$ moment in the undoped system. This continuous reduction reflects the oxidation of Ni$^{2+}$ ($d^8$, $e_g^2$, $S=1$) toward Ni$^{3+}$ ($d^7$, $e_g^1$), which can favor low-spin ($S=1/2$) or even nonmagnetic states depending on local environments. The emergence of a nonmagnetic Ni site at $x=1.0$ strongly supports the picture of charge and orbital disproportionations.
These findings highlight the intricate role of Sr doping in tuning the spin state and local electronic structure of Ni ions, offering a route to modulate the magnetic and transport properties of layered nickelates.

{\bf Charge and Orbital Ordering in LaSrNiO$_4$:}
Up to date, La$_3$Ni$_2$O$_7$ has been experimentally observed that only spin ordering persists, with no detectable signatures of charge or orbital ordering \cite{Zhao2025-ScienceBulletin,Shi2025-arxiv}. However, multiple theoretical investigations have consistently suggested the coexistence of coupled spin-charge orders in its magnetic ground state \cite{Ni2025-NPJ,Zhang2024-arxiv,LaBollita2024-PRM}.
These results align with experimental evidence of magnetic moment differentiation between distinct Ni sites \cite{Zhao2025-ScienceBulletin,Shi2025-arxiv}, as magnetic ordering with notable spiltting of magnetic moments may naturally lead to concomitant weak charge ordering and  weak the orbital ordering, as shown in Table \ref{T2}.
\begin{table}[htbp]
  \setlength{\belowcaptionskip}{15pt}
  \caption{\label{T2} The charge, spin and orbital occupations of two inequivalent Ni atoms in LaSrNiO$_4$ ($x$ = 1).}
  \renewcommand{\arraystretch}{1.2}
  \begin{ruledtabular}
  \begin{tabular}{c|ccccc}
  Atom (spin) & $d_{xy}$ & $d_{yz}$ & $d_{xz}$ & $d_{x^{2}-y^{2}}$ & $d_{3z^{2}}$ \\
  \hline
  Ni-1 (up)      & 0.98 & 0.98 & 0.98 & 0.71 & 0.67 \\
  Ni-1 (down)    & 0.98 & 0.98 & 0.98 & 0.71 & 0.67 \\
  Ni-2 (up)      & 0.95 & 0.96 & 0.96 & 0.42 & 0.40 \\
  Ni-2 (down)    & 0.98 & 0.98 & 0.98 & 0.92 & 0.94 \\
  \end{tabular}
  \end{ruledtabular}
\end{table}
From this table, we find that in addition to the magnetic moment difference between Ni-1 and Ni-2, there are weak charge order with $n_{\text{Ni-1}}-n_{\text{Ni-2}}$ = 0.08, and weak orbital order with $n_{x^{2}-y^{2}}-n_{3z^{2}}$ = 0.08 on atom Ni-1.
To further clarify the charge transfer and oxidation states, we performed a Bader charge analysis and virtual crystal approximation (VCA), the details are available in Section III and IV of the Supplemental Material. Consequently, the coexistence scenario of spin order, weak orbital and charge orders in LaSrNiO$_4$ is highly plausible. We expect further experimental verification.

{\bf Discussions and Conclusions:}
On the same time, though single Ni ion in La$_2$NiO$_4$ has the similar crystal field environment to those in La$_3$Ni$_2$O$_7$ and La$_4$Ni$_3$O$_{10}$, the Ni spins of the NiO$_2$ plane in 214 phase display quasi-two-dimensional characteristic. Thus, our strong correlation correction with $U$ = 5 eV to the electronic structures is justified and reasonable.
Furthermore, to some extent electron or hole doping plays similar role with the varying hydrostatic pressure, i.e. the chemical pressure. For example, Ca, Sr or other bivalent elements doping not only introduces hole carriers, but also alters the orbital occupations, as well as the band structures and Fermi surfaces. We find that the doping plays distinct role on magnetic configurations, in comparison with  the variation of the hydrostatic pressure in La$_2$NiO$_4$.

The contrasting behavior between La$_2$NiO$_4$ and La$_3$Ni$_2$O$_7$ nickelates highlights the critical role of dimensionality in determining electronic correlations and magnetic interactions. While La$_3$Ni$_2$O$_7$ benefits from pressure-enhanced interlayer $d_{z^2}$ hybridization that weakens magnetism and enables superconductivity, La$_2$NiO$_4$ lacks this mechanism due to its isolated NiO$_2$ planes. The robust magnetism observed under both pressure and doping suggests that achieving superconductivity in the 214 phase may require more sophisticated approaches, such as strain engineering, interfacial effects, or heterostructuring.

In summary, we have shown that in the ambient pressure tetragonal La$_2$NiO$_4$ exhibits Neel-type antiferromagnetic order in planar Ni spins with negligible interlayer magnetic coupling; with the increase of pressure, La$_2$NiO$_4$ evolves from insulate into metallic state with the critical pressure when P is larger than 50 GPa; with the substitution of La by Sr, the magnetic ground state evolves from G-type to double spin striped antiferromagnetic one, as well as weak charge and orbital orders at $x=1$.
These results not only favor to clarify the evolution of the groundstate magnetic structure, charge distributions and orbital structures of the parent phase La$_2$NiO$_4$ with the increasing pressure, but also are benefitful for our getting insight of the magnetic fluctuations and the superconductive pairing mechanism of the Ruddlesden-Popper compounds.

\begin{acknowledgements}

This work is supported by the National Natural Science Foundation of China (NSFC) under Grant Nos. 11974354, U2030114 and the CASHIPS Director Fund under Grant No. YZJJ202207-CX. Numerical  calculations were partly performed in the Center for Computational Science of CASHIPS, the ScGrid of Supercomputing Center, Computer Network Information Center of Chinese Academy of Sciences, and the Hefei Advanced Computing Center.

\end{acknowledgements}

\bibliography{La2NiO4}

\end{document}